\documentclass[sigconf, nonacm]{acmart}

\usepackage[ruled,linesnumbered]{algorithm2e}
\usepackage{hyperref}
\usepackage{enumitem}
\usepackage{xspace}

\usepackage{amssymb}
\usepackage{framed}
\usepackage{color}
\usepackage[most]{tcolorbox}
\usepackage{multirow}
\usepackage{colortbl}
\usepackage{makecell} 
\usepackage{booktabs} 
\usepackage{bbding}

\newcommand{\bench}{\textbf{\textit{DQABench}}\xspace}
\newcommand{\benchit}{\textit{DQABench}\xspace}
\newcommand{\testbed}{\textit{DQATestbed}}
\newcommand{\qa}{QA\xspace}

\newcommand{\grey}{\cellcolor{lightgray}}
\newcommand{\hl}[1]{\textcolor{purple}{#1}}

\definecolor{darkblue}{rgb}{0, 0, 0.5}
\hypersetup{colorlinks=true, citecolor=darkblue, linkcolor=darkblue, urlcolor=darkblue}
\definecolor{lightgray}{rgb}{0.9, 0.9, 0.9}

\definecolor{darkgreen}{RGB}{50,100,0}
\definecolor{darkred}{RGB}{200, 0, 0}
\definecolor{lightred}{RGB}{250, 200, 200}
\definecolor{lightblue}{RGB}{210, 220, 250}
\definecolor{doderblue}{RGB}{30,144,255}
\definecolor{select}{RGB}{222, 235, 247}
\definecolor{unselect}{RGB}{247, 207, 206}

\setlist[itemize]{left=1cm, right=1cm, topsep=1cm, partopsep=1cm}

\begin{document}
\title{Revolutionizing Database Q\&A with Large Language Models: Comprehensive Benchmark and Evaluation [EA\&B]}


\author{Yihang Zheng¹, Bo Li¹, Zhenghao Lin¹, Yi Luo¹, Xuanhe Zhou², }
\author{Chen Lin¹*, Jinsong Su¹, Guoliang Li², Shifu Li³}
\author{¹ Xiamen University; ² Tsinghua University; ³ Huawei}

\begin{abstract}
The development of Large Language Models (LLMs) has revolutionized \qa across various industries, including the database domain. However, there lacks a thorough evaluation regarding the capabilities of different LLMs in database \qa. To this end, we introduce \benchit, the first comprehensive database \qa benchmark for LLMs. \benchit features an innovative LLM-based method to automate the generation, cleaning, and rewriting of evaluation dataset, resulting in over 200,000 \qa pairs in English and Chinese. These \qa pairs cover a wide range of database-specific knowledge extracted from manuals, online communities, and DB instances, allowing for assessment of LLMs' Retrieval-Augmented Generation (RAG) and Tool Invocation Generation (TIG) capabilities in the database \qa task. Furthermore, we propose a highly modular and scalable testbed \testbed, with basic and advanced components such as Question Classification Routing (QCR), RAG, TIG, and Prompt Template Engineering (PTE). Finally, we provide an evaluation pipeline that computes various metrics throughout a standardized evaluation process to ensure the accuracy and fairness of the evaluation. Our evaluation reveals the strengths and limitations of nine open-source and commercial LLMs and the impact of various service components (e.g., QCR, RAG, TIG). The proposed benchmark dataset, testbed, and findings will guide the future development of LLM-based database applications.
\end{abstract}

\pagestyle{plain}
\pagenumbering{arabic}

\maketitle

\section{Introduction}

Large language models (LLMs) have emerged as one of the most promising artificial intelligence technologies in recent years. LLMs have achieved remarkable progress in answering diverse questions from vast domains, such as medicine~\cite{medicine1,medicine2}, finance~\cite{finance}, earth science~\cite{earth}, law~\cite{law}, and so on. 

\begin{figure*}[!t]
  \centering
 \includegraphics[width=\linewidth]{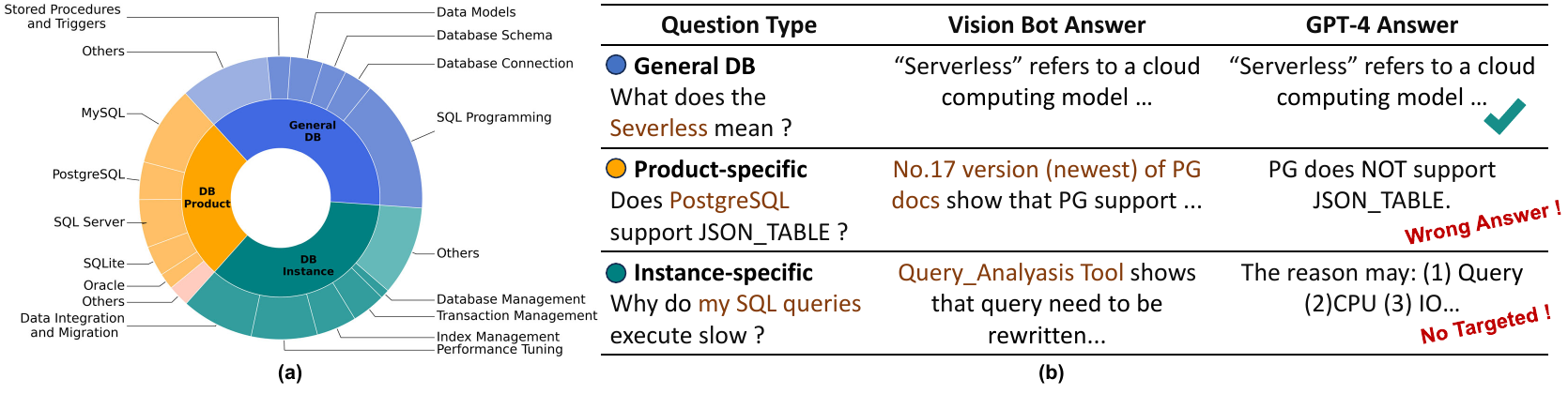}
  \vspace{-1em}
  \caption{(a) Percentage of various questions in online DB communities. (b) \qa examples (GPT-4 v.s. Ground-truth) }
  \label{fig:intro}
  \vspace{-1em}
\end{figure*}

Question-answering (QA) systems are pivotal copilots for database systems. Traditionally, extensive IT knowledge, such as SQL, storage, hardware, and network, is necessary for database deployment and usage. Advanced operations, such as configuration and optimization, require skilled database administrators (DBAs) to handle effectively. This demand for expertise highlights the necessity for Database Question Answering (DBQA), which allows DB users to pose questions freely and receive technical support whenever and wherever needed. DBQA will empower businesses to manage DBMS quickly and efficiently while reducing labor costs.

People would anticipate that LLMs also excel in DBQA. However, \textbf{the potential of LLMs in DBQA remains unexplored}. As illustrated in Figure~\ref{fig:intro} (a)\footnote{The distribution is calculated based on $68,801$ questions collected from StackExchange.}, database questions can be generally categorized into three types. Although LLM evaluation has attracted a lot of attention from natural language processing~\cite{gpt4, llama} and database communities~\cite{text2sql,text2sql2,datamanagement}, \textit{existing evaluations can not fully reflect LLM's ability to exploit diverse DB-specific reasoning channels to generate answers}, as specified in the following.

(1) \textit{General DB questions} involve fundamental concepts in the database domain, e.g., SQL grammar, E-R model, etc. Answering general DB questions requires the LLM to encapsulate substantial database knowledge. Although a few QA benchmarks~\cite{wiki,MMLU} contain IT-related questions, they do not cover advanced or up-to-date database concepts such as ``serverless'' and ``object storage''. (2) \textit{Product-specific questions}, which relate to using particular database products, e.g., setting up a local PostgreSQL database, operational instruction for Snowflake, etc. Answering product-specific questions requires the LLM to consult external sources such as product manuals or enterprise instructions because of data privacy or information timeliness issues. Existing Retriever-Augmented Generation (RAG) benchmarks~\cite{ragbench,ragbench2,wiki} predominantly assess retrieval from news and wiki sources. While database documents are more technical (e.g., long contexts, professional concepts, rigorous structures with code blocks, etc.), retrieving textual pieces that directly address the question to enhance the answer is more complex. (3) \textit{Instance-specific questions} centering on a particular database instance, e.g., fault diagnosis and data analysis for a bank's database running on Postgres v16 with Intel i7 CPU. Answering instance-specific questions depends on context information provided by DB tools, e.g., statistics from various system views in the database. Current Text2SQL benchmarks~\cite{spider,text2sql,text2sql2} emphasize only SQL generation capabilities, while LLM-based agent benchmarks~\cite{toolformer} focus on the simple invocation of a broad pool of tools. In contrast, the database tools have strict format requirements for the input instead of SQL queries, posing challenges for the LLM's instruction-following capabilities. Moreover, multi-turn invocation is often necessary, demanding tool planning abilities for LLMs.

As shown in Figure~\ref{fig:intro}(b) \footnote{Showcases are abbreviated from answers generated by GPT-4.}, even the most powerful commercial LLM (i.e., GPT-4) cannot always give correct answers. The limitation of GPT-4 emphasizes the importance of a comprehensive evaluation that can enable us to understand the challenges and opportunities of DBQA in the era of LLMs. Conducting such an evaluation faces the following challenges.

\textbf{C1: DBQA Dataset}. Current studies are mostly based on data collected from the Web~\cite{SQuAD,wiki,amazon,reddit} instead of manual composition to increase scalability, raising three issues in the DB domain. (1) \textit{Low Question Quality}. Online questions are too brief and lack essential contextual information. For example, a question on slow database performance without hardware (CPU, IO) or query information is impossible to diagnose. (2) \textit{Low Answer Quality}. Many online answers are factually incorrect, overly concise, or subjective. (3) \textit{Limited Diversity}. Due to factors such as conformity bias, questions in online communities tend to center on a narrow range of topics and DBMS products, e.g., users often hesitate to ask "silly" questions, and most questions concern popular DBMS products.

\textbf{C2: DBQA Testbed}. Previous evaluations of LLM~\cite{datasetSurvey} primarily focus on a standalone LLM. Unfortunately, regardless of the LLM backbone architectures, a series of components are indispensable in DBQA, including: \textit{pre-training} to equip the LLM with domain knowledge to answer DB general questions, \textit{fine-tuning} to enhance the LLM to follow DB-specific instructions, \textit{routing} to adopt different reasoning logics for various types of questions, \textit{retrieving} to consult an external knowledge source to answer product-specific questions, and \textit{tool invocating} to interact with the DB environment to answer instance-specific questions. Thus, a testbed is required to support all LLMs and integrate these components to investigate various functionalities related to DBQA.

\textbf{C3: DBQA Evaluation}. Existing benchmarks~\cite{medicine1,medicine2} fail to compare LLMs in aspects crucial in the database domain, such as the factual accuracy and technical depth of the answers, rather than the fluency of explanation. Moreover, they focus on the final answers and lack reasonable metrics to measure the fine-grained modular (e.g., intermediate components) performance and end-to-end performance (e.g., eliminate the impact of intermediate components). 

To address these challenges, we construct \bench (Database Question-Answer benchmark) and present a thorough evaluation based on \bench. \textbf{To address C1}, we build a \qa dataset by enriching queries from online DB resources and enhancing answer quality through cleaning and rewriting. We also apply a few-shot learning and chain-of-thought approach to extract \qa pairs from DBMS documents and instances, increasing dataset diversity. \textbf{To address C2}, we propose a testbed that incorporates all components of a complete DBQA bot, including pre-training, fine-tuning, Question-Category Routing (QCR), Prompt-Template Engineering (PTE), RAG, and Tool-Invocation Generation (TIG), each optimized to adapt general-purpose LLMs for DBQA tasks. \textbf{To address C3}, we implement a standardized end-to-end evaluation pipeline, which reduces uncertainty and increases evaluation fairness in comparing LLMs' answers. Additionally, we develop a modular evaluation framework to assess the performance of different solutions in intermediate components such as QCR, RAG, and TIG.

In summary, we make the following contributions. 

\begin{sloppypar}
(1) We propose the first benchmark dataset \bench to evaluate question-answering performance in the database domain. \bench simulates real-world DB scenarios and covers questions regarding DB general knowledge and complex questions that demand assistance from external manuals and DB tools. The dataset contains 200,000 \qa pairs, larger than existing instruction datasets in the IT field~\cite{datasetSurvey}. (Section \ref{sec:bench})
\end{sloppypar}

(2) We propose a plug-and-play testbed to experiment with different LLM application strategies. The testbed assembles all components potentially involved in DBQA, such as QCR, PTE, RAG and TIG.  (Section \ref{sec:framework})

(3) We conduct an in-depth evaluation of the end-to-end performance of nine LLMs (Section \ref{sec:end2end}) and their modular performance using various components, including different RAG solutions and classifiers for question type categorization. We discover several insights, including but not limited to the following five key aspects. (Section \ref{sec: modular})

\noindent \textbf{I1: Performance disparity of LLMs}. LLMs present significant variation in their ability to answer database questions. Larger model sizes generally lead to better performance, while small models are limited in handling advanced questions that involve tool usage and understanding lengthy knowledge sources. 

\noindent \textbf{I2: Importance of pre-training and fine-tuning}. Pre-training plays a crucial role in developing a broad understanding of database-related knowledge, whereas fine-tuning significantly enhances performance across all tasks, especially those requiring instruction-following abilities, such as tool invocation. We find that with proper pre-training and fine-tuning, even small-sized open-source LLMs can outperform openAI's GPT-4. 

\noindent \textbf{I3: Necessity of question routing}. We verify that question routing is crucial because complex DB questions cannot be resolved solely through the internal reasoning of a standalone LLM model. Instead, different question types must trigger special treatments in the auxiliary components such as RAG or LLM agents. Moreover, we show that routing can not rely on LLM's response but needs a well-trained classifier.

\noindent \textbf{I4: Impact of knowledge retrieval}. We demonstrate the potential of RAG in enhancing the LLM's response under the condition that relevant knowledge is accurately obtained. Nonetheless, we reveal the primary bottleneck lies in the low recall rate, i.e., the inability to identify relevant knowledge, which remains an open challenge for future DBQA.

\noindent \textbf{I5: Lack of tool selection and tool invocation abilities}. We demonstrate the incompetence of open-source LLMs in answering instance-specific questions because they cannot select and invoke appropriate DB tools. Fine-tuning the model's instruction-following abilities can enhance LLM agents and improve DBQA performances.

\section{Related Work}
\subsection{\qa by Large Language Models}

\noindent \textbf{\qa with General-purpose LLMs}. Researchers have discovered that model performance can be enhanced by scaling up the model size and training data. Large-scale LLMs such as GPT-3.5~\cite{chatgpt}, LLaMa~\cite{llama}, and PaLM~\cite{palm} have emerged. These models demonstrate \qa capabilities far beyond traditional pre-trained language models. For instance, GPT-4~\cite{gpt4} achieves human-level performance on most professional and academic examinations. Medium-sized LLMs have also been developed to meet the demand for edge deployment. Examples include Llama3-7B~\cite{llama}, Mistral-7B~\cite{mistral}, Baichuan2-13B~\cite{baichuan} and Qwen-14B~\cite{qwen}. Recently, small models, such as Yuan-2B~\cite{yuan}, have been shown to approximate the performance of medium-sized LLMs. However, research indicates that the performance of medium-sized and small models is constrained to simple, singular tasks.

\noindent \textbf{\qa with Domain-Specific LLMs.} In vertical domains such as medicine~\cite{medicine1,medicine2}, finance~\cite{finance}, earth science~\cite{earth} and law~\cite{law}, due to (1) the diverse language styles of questions and (2) the complexity and depth of expertise, many customized LLMs have been developed. These LLMs typically underwent domain-specific pre-training and fine-tuning. Consequently, many smaller-scale models can achieve or even surpass the \qa abilities of GPT-4 in their respective domains.

\noindent \textbf{\qa with knowledge-augmented LLMs}. To ensure the quality of \qa, researchers have introduced external knowledge to enhance answer generation. The external knowledge can be brought as follows: (1) by retrieving documents~\cite{rag} or guidelines\cite{luo_safety} for questions to ground the answers; (2) by using a structured knowledge base, such as knowledge graphs, for reliable reasoning~\cite{tog}; (3) by using LLMs as an agent to trigger external tools~\cite{agent} to solve specific sub-tasks or obtain contextual information.

\subsection{Large Language Models for Database}

\noindent \textbf{LLMs for Database Management}. Since LLMs have demonstrated outstanding capabilities in knowledge comprehension and contextual understanding, researchers in the database domain have started to explore LLMs for various database-related tasks. Raul et al.~\cite{vldbdata} argue that LLMs can ground database tuples, schemas, and queries in novel ways. Zhou et al.~\cite{llmasdba} propose that LLMs can serve as Database Administrators (DBAs).  D-Bot~\cite{dbot} applies LLMs for intelligent database diagnosis. Liu et al.~\cite{queryrewriting} use LLMs for query rewriting. However, these studies focus on specific database management tasks rather than answering real-life user questions.

\noindent \textbf{LLMs for NL2SQL.} LLMs have shown impressive capabilities in converting natural language into SQL queries, bringing about significant changes in simplifying user interactions with databases. Works such as Binder-SQL~\cite{binsql}, DIN-SQL~\cite{dinsql}, and BIRD~\cite{li2024can} enable LLMs to generate corresponding SQL statements directly from input objectives in natural language. DB-GPT~\cite{db-gpt-xue} allows users to input their requirements in natural language and receive complete visualized data analysis and reports. In addition, some benchmarks are proposed to evaluate LLMs' performances on NL2SQL~\cite{text2sql,text2sql2}. Compared with NL2SQL, our paper considers tool invocation by LLMs, which is not limited to SQL query generation because the LLM's generation must accord with the format requirements of various tools. 

\section{\bench Dataset Generation}\label{sec:bench}

\begin{table*}
\centering
\caption{\bench dataset statistics}\label{tab:dataset}
\vspace{-.5em}
\begin{tabular}{c|c|r|r|r|r|r|c} 
\hline
\multirow{2}{*}{\textbf{Type}}              & \multirow{2}{*}{\textbf{Source}} & \multicolumn{1}{c|}{\multirow{2}{*}{\textbf{\# Q.}}} & \multicolumn{2}{c|}{\textbf{English}}                                          & \multicolumn{2}{c|}{\textbf{Chinese}}                                          & \multirow{2}{*}{\textbf{Annotation}}       \\ 
\cline{4-7}
                                   &                         & \multicolumn{1}{c|}{}                       & \multicolumn{1}{l|}{\textbf{Avg. Q. Len.}} & \multicolumn{1}{l|}{\textbf{Avg. A. len.}} & \multicolumn{1}{l|}{\textbf{Avg. Q. Len.}} & \multicolumn{1}{l|}{\textbf{Avg. A. len.}} &                                   \\ 
\hline
\multirow{2}{*}{\textbf{General}}           & Exam                    & 2,152                                       & 224.19                                  & 458.61                                 & 85.44                                  &  124.59                                 & \multirow{2}{*}{N/A}              \\ 
\cline{2-7}
                                   & Forum  & 74,893  & 1205.75             & 1273.16             & 354.15          & 790.24          &                 \\ 
\hline
\multirow{2}{*}{\textbf{Product-Specific}}  & OpenGauss       & 21,689 & 78.19               & 666.26              & 31.34           & 267.42                                             & \multirow{2}{*}{Retrieval Label}  \\ 
\cline{2-7}
                                   & GaussDB         & 4,950  & 84.84               & 885.67              & 34.62           & 394.60          &                 \\ 
\hline
\multirow{2}{*}{\textbf{Instance-Specific}} & Common           & 1,080  & 86.07               & 1070.21             & 29.76           & 757.23                                            & \multirow{2}{*}{Tool Label}       \\ 
\cline{2-7}
                                   & Generalization   & 2,707  & 127.18              & 1019.28             & 34.81           & 515.92         &                                   \\
\hline
\end{tabular}
\vspace{-.5em}
\end{table*}

A dataset consisting of pairs of questions and answers is crucial for evaluating the performance of a DataBase Question-Answering (DBQA) bot. However, manually creating such pairs is labor-intensive. Publicly accessible data often fails to encompass the diverse topics that database users might propose. This section introduces techniques for generating a dataset tailored to the DBQA benchmark.

As shown in Figure~\ref{fig:intro} (a), the DB questions can be divided into three subsets, corresponding to three key skill sets of an LLM-based DBQA bot. \textit{General DB questions} can evaluate whether the bot grasps DB-related concepts and knowledge. \textit{Product-specific questions} can evaluate whether the bot applies knowledge of the target database product. \textit{Instance-specific questions} can evaluate whether the bot adapts to real-life DB circumstances. 

The three question categories are different in (1) data sources: general DB questions are publicly available, while product-specific questions and instance-specific questions are almost impossible to obtain complete examples online. (2) problem background: the latter two categories (i.e., product-specific questions and instance-specific questions) need supporting information, such as the product manual and instance context. (3) ground-truth answers: the latter two categories must provide retrieval results or tool invocation results to demonstrate the reasoning and produce trustworthy answers.

\textbf{Data Statistics}. Accordingly, we propose methods to construct the dataset for each category. As shown in Table \ref{tab:dataset}, we construct a dataset with bi-lingual \qa pairs on the three categories, translating English pairs into Chinese and vice-versa, leading to a total of over 200,000 \qa pairs. 

\vspace{-.5em}
\subsection{General DB \qa}\label{sec:bench_general}
\begin{table}
    \centering
    \caption{Example of general DB \qa}
    \vspace{-1em}
    \begin{small}
    \begin{tabular}{p{1.5cm}p{6cm}}
    \toprule
    \textbf{Question} &
    How to create ASM and install Grid for beginners confused about hard drives and disks?
        \\\midrule
    \textbf{Original Answer} &
    If you're not clear on hard drives or disks, start with basic Linux knowledge first.
        \\\midrule
    \textbf{Prompt} &
    ... convert it to a ``detailed, professional and friendly" writing style.
        \\\midrule
    \textbf{Rewritten Answer} &
    ..., I \textit{recommend} starting with some basic Linux knowledge, e.g., \textbf{Device Management: fdisk, lsblk, df} …. Firstly, ASM is a storage management solution… Typically, ASM requires \textbf{at least two disks to create a disk}… Before creating ASM, you should use…\\
    \bottomrule
    \end{tabular}
    \label{tab:generalbench}
    \end{small}
    \parbox{0.95\linewidth}{\small \textbf{Note:} \textbf{Bold} - extended knowledge; \textit{Italic} - friendly tone.}
\vspace{-.5em}

\end{table}

\textbf{Data Sources}. We have two types of data sources. (1) Similar to other domain-specific datasets, exams are a good data source because exams offer objective and typically accurate ground-truths. We collect $2,000$ unique multiple-choice questions from four DB university textbooks, three online courses, and 28 online course exams. (2) To ensure that the \bench dataset covers a wide range of questions asked by DB users in daily usage, we collect \qa entries from the largest English and Chinese online DB communities, namely the database section of StackOverflow\footnote{\url{https://stackoverflow.com/questions/tagged/database}}, the Database Administrators section of StackExchange\footnote{\url{https://dba.stackexchange.com/}}, and MoDB\footnote{\url{https://www.modb.pro/}}.

\textbf{Step 1: Question Filtering}. One major problem of online \qa pairs is that the answers are not guaranteed to be factually accurate. Thus, we filter the collected content based on online feedback. First, we compute the ROUGE-1 score, which measures the overlap of unigrams between questions. \qa with a large ROUGE-1 score ($\geq 0.8$) are merged to de-duplicate the questions and reduce possible grammatical problems. For each question, we retain only the accepted answers and those with high upvotes to ensure the factual correctness of the responses.

\textbf{Step 2: Answer Rewriting}. The answers collected are insufficient as ground-truth answers. For example, the exam questions are only associated with letter options, and the LLMs' generation may be too random when only generating a letter option. Therefore, for each exam question, we extend the answer by instructing GPT-4 to provide detailed explanations for the answer choices. Meanwhile, online responses are also often overly concise, emotional, and subjective. As shown in Table~\ref{tab:generalbench}, while the replies explicitly advise the inquirer to learn basic Linux knowledge, they do not specify a learning path or key points, and the tone is not user-friendly. For each online question, we reform the accepted response by instructing GPT-4 to convert it to a ``detailed, professional and friendly" writing style. Table \ref{tab:generalbench} shows the prompts, and the rewritten results are more specific and friendly.

\subsection{Product-Specific \qa}\label{sec:bench_product}
\begin{figure}[!t]
  \centering
\includegraphics[width=\linewidth]{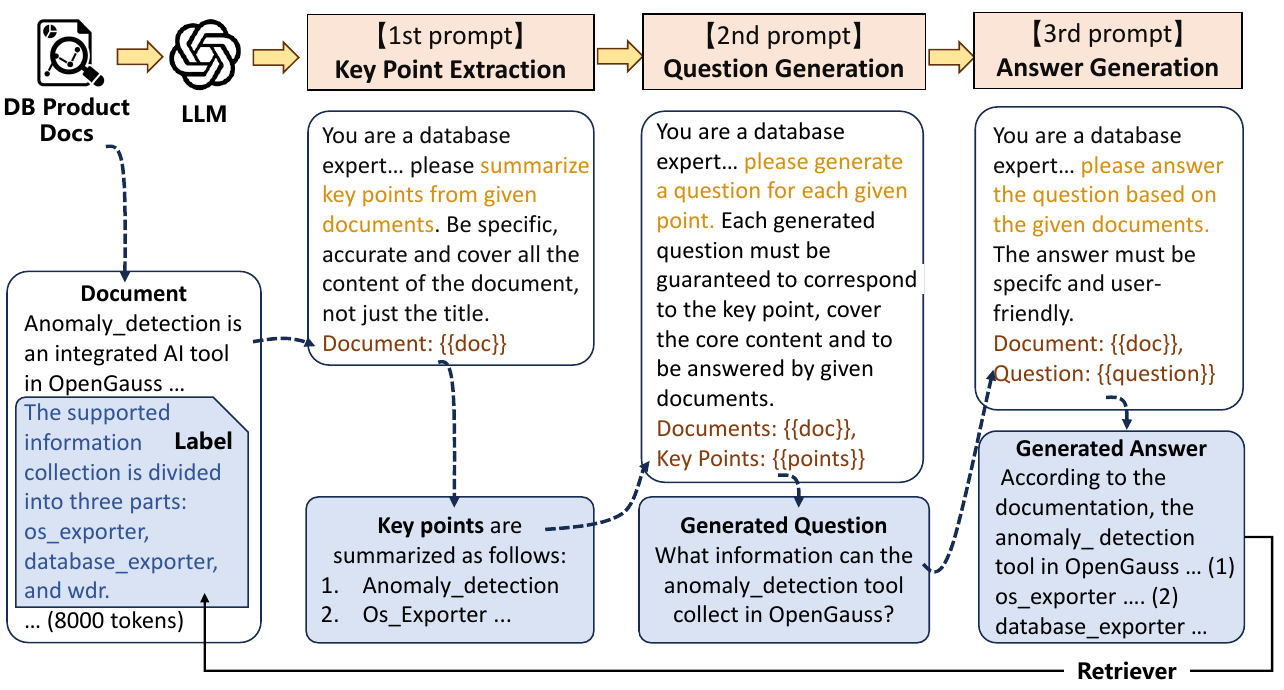}
  \caption{Generation of product-specific \qa}
  \label{fig:productbench}
  \vspace{-1em}
\end{figure}
Constructing product-specific \qa pairs from online sources can be problematic because it is impossible to tell whether the online answers are based on a particular product documentation and even locate the product documentation as a ground-truth for evaluation. Thus, we construct the product-specific \qa pairs via the workflow illustrated in Figure~\ref{fig:productbench}. 

\textbf{Step 1: Pre-processing Manuals}. Most product manuals are too lengthy for LLMs to comprehend. Thus, we pre-process each product manual, i.e., we segment the documents, where each segment contains complete paragraphs while not exceeding 8,000 tokens. This segmentation allows LLMs to process the documents at a finer granularity, thereby generating more detailed and comprehensive \qa. 

\textbf{Step 2: \qa Generating}. To reduce manual efforts, we use LLMs to generate several \qa pairs on each document segment. The challenge is, directly instructing the LLM to generate \qa results in low-quality outcomes. Specifically, the generated questions can overly focus on minor details while neglecting the main points of the given document segment, the answers can be too concise, and the QA pairs can be repetitive, lacking a diverse coverage of possible topics. Thus, we propose a novel prompt chain to generate \qa. As shown in Figure~\ref{fig:productbench}, the prompt chain first requires LLMs to summarize the document segment's key points. Then, the prompt chain demands LLMs to generate a question for each key point that can be answered based on the document segment. Finally, the prompt chain asks LLMs to produce a detailed, user-friendly answer.

\textbf{Step 3: Retrieval Label Annotating.} As this dataset supports evaluating the \qa bot's intelligence in applying external knowledge and adapting to different DB products in the RAG (Retrieval Augmented Generation) scenarios, in addition to providing \qa pairs, we also annotate the relevant text chunk. To more precisely locate a finer-grained passage instead of the whole document, we store the text chunks in each document segment in the vector database, using the generated question and answer as a query, and pinpoint the set of text chunks containing relevant information (with cosine similarity $\geq$ 0.8).

\begin{table}[!t]
 \caption{Supported types of common DB tools}\label{tab:tool}
 \begin{small}
\begin{tabular}{p{1.5cm}|p{1cm}|p{5cm}}
\hline
\textbf{Objective}                            & \textbf{Type}      & \textbf{Functionality}                                                                            \\ \hline
\multirow{2}{1.5cm}{\textbf{\\ Data \\ Modeling}}       & Schema    & Obtain database table structure, constraints, etc.                                       \\ \cline{2-3} 
                                     & Selection & Return SQL execution results of retrieving specific data from the database, computing data distribution, etc. \\ \hline
\multirow{3}{1.5cm}{\textbf{\\ Database\\ Monitoring}} & Resource   & Obtain information about CPU usage, memory, disk IO, etc.                                \\ \cline{2-3} 
                                     & Workload  & Workload analysis, slow question identification, etc.                                       \\ \cline{2-3} 
                                     & Status    & Detailed information about the current indexes, views, knob settings, etc.                     \\ \hline
\textbf{Optimizing}             & Tuning    & Identify optimization opportunities by advising indexes, setting knobs, etc.                        \\ \hline
\end{tabular}
\end{small}
\end{table}

\subsection{Instance-Specific \qa}\label{sec:bench_instance}

It is infeasible to construct instance-specific \qa pairs from online sources. Online questions are almost always incomplete due to privacy reasons, missing necessary instance-level contextual information, such as the database's table structures, workload information, etc. It is impractical to conduct DB interaction with the specified DB instance referred to in the online question to restore the contextual information. Therefore, we have to generate instance-specific \qa pairs by LLMs automatically. 

\begin{figure}[!t]
  \centering
\includegraphics[width=\linewidth]{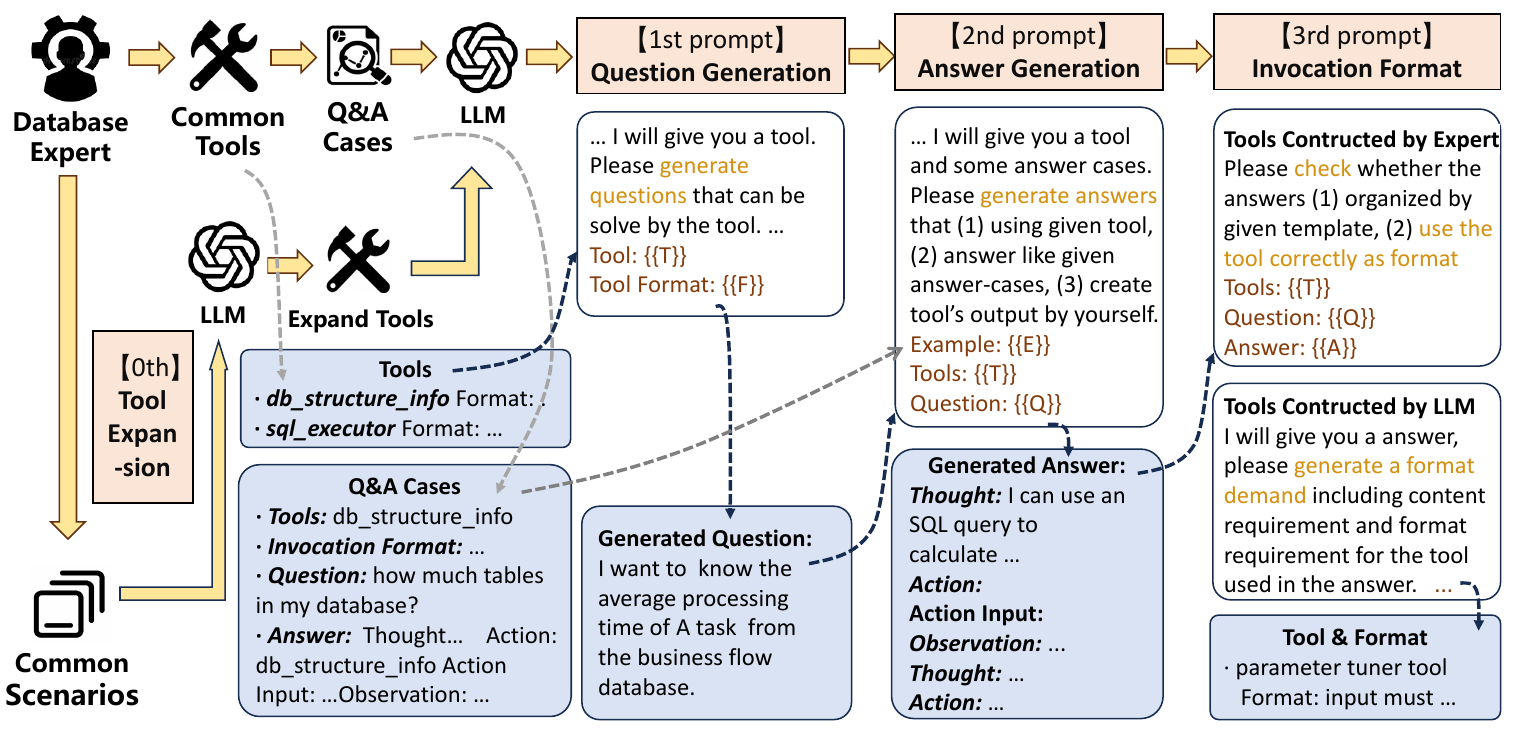}
  \vspace{-1em}
  \caption{Generation of instance-specific \qa}
  \label{fig:instancebench}
  \vspace{-2em}
\end{figure}

There are numerous database \textit{tools} provided in DBMS that support database monitoring, optimizing, and analyzing for DB developers, administrators, and analysts. The LLM's proficiency in answering instance-specific questions relies on whether LLMs can accurately invoke different tools to obtain the instance's contextual information. Thus, as shown in Figure~\ref{fig:instancebench}, our dataset construction workflow starts with building a pool of DB tools.

\textbf{Step 1: Constructing DB Tool Pool}. 
(1) We first survey the DB tools commonly used in real-production systems. As shown in Table~\ref{tab:tool}, we identify six types of \textit{common DB tools} that are frequently used for data modeling, database monitoring and performance optimization. The implementation of each tool type, including the exact tool name and the format of input and output, may vary for different DBMS products.\footnote{Implementation Details on PostgreSQL and OpenGauss are shown in \href{https://github.com/XMUDM/DQABench}{[link]}} (2) The common tools can not cover all DB tools, especially new ones developed to meet the demands of a real-production system. We expand the common tools to a set of \textit{generalization tools} to evaluate the \qa bot's generalization ability to utilize different DB tools properly. We include scenarios such as Operation Diagnostics, Business Intelligence, Performance Monitoring and Tuning, Data Analysis, System Utilization Analysis, Deployment Issues, Operations Management, question Optimization, Backup and Recovery, Permission Management, Index Management, and Database Tuning. We require GPT-4 to generate imaginary DB tools that can benefit the above DB scenarios. 

\textbf{Step 2: Generating Questions}. 
For each tool, including common tools and generalization tools, we require GPT-4 to generate questions that can be solved by the target tool using the prompt in Figure~\ref{fig:instancebench}. 
Moreover, we want to effectively evaluate the \qa bot's \textit{planning} ability, which involves adequately combining several DB tools and organizing tools with the right action order. Thus, we manually construct three to four questions for each common tool and scenario that demand a chain of multiple tool invocations, and we use these questions as few-shot examples to encourage GPT-4 to generate complex questions. We post-process the resulting questions to ensure that more than $50\%$ of the generated questions are solved by invoking at least two DB tools. 

\textbf{Step 3: Generating Answers}. (1) First, we manually produce \textit{answer cases} for manually constructed questions above. The DB experts compose the answer cases in the following procedure: construct a real DB instance according to the description in the question, call the DB tools when necessary, and answer the question based on real tool feedback. (2) Then we use the answer cases as few-shot learning examples to guide GPT-4 to generate answers efficiently. We adopt the Chain-Of-Thought(COT) prompting technique to generate an answer for each question. 
The prompt~\cite{react} encourages GPT-4 to break down the problem of answer generation into a series of intermediate reasoning steps. Each step corresponds to either tool invocation or answer generation,  including ``\textit{Thought}", ``\textit{Action}," ``\textit{Action\_Input}," ``\textit{Observation}," and ``\textit{Answer}." Here, ``\textit{Thought}" is the logical reasoning, ``\textit{Action}" and ``\textit{Action\_Input}" are the tool's name and the tool's input, ``\textit{Observation}" is the instance's information returned by calling the tool. This way, even if errors occur, e.g., GPT-4 produces incorrect actions to trigger the tool or the observation does not simulate the tool's output, GPT-4 can switch to alternative approaches, resulting in more accurate and reliable answers. 

\textbf{Step 4: Polishing Answers}. Finally, we ask GPT-4 to rethink and polish the answer. This step is different for common tools and generalization tools. (1) For each answer to common tools, since the common tools are real DB tools with pre-defined formats of tool output, we ask GPT-4 to examine its output to ensure the answer format is correct to trigger the tool. (2) For each answer relating to generalization tools, since the generalization tools are imagined and reasonably inferred by GPT-4, they do not have a pre-defined format; we ask GPT-4 to summarize the tool's format.

\section{\testbed} \label{sec:framework}

\begin{figure*}[!t]
\vspace{-1em}
  \centering
\includegraphics[width=0.95\linewidth]{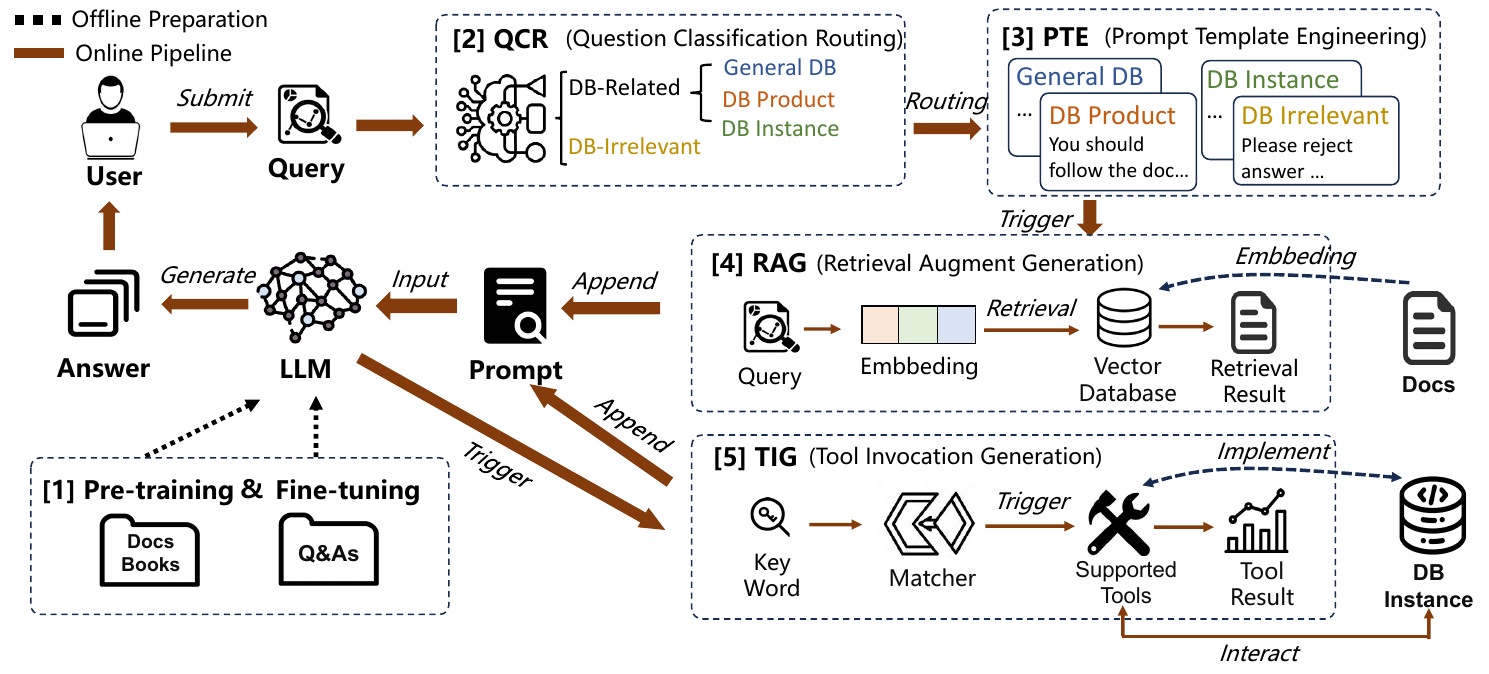}
  \vspace{-1em}
  \caption{Framework overview of \testbed}
  \label{fig:system}
  \vspace{-1em}
\end{figure*}

When adopting a general-purpose LLM for DB \qa, various auxiliary modules are indispensable to leverage and adapt the LLM's general knowledge of linguistic patterns and common senses into the DB environment. Currently, there lacks an all-encompassing DBQA testbed that incorporates LLM and various auxiliary modules. Table \ref{tab:framework} compares the completeness of the proposed testbed with recent LLM adaptation frameworks in the open domain and DB domain. Existing works overlook some important modules. 
On the contrary, our proposed testbed supports a full chain of auxiliary modules for LLM's domain adaptation. We believe that these modules represent a future design paradigm for database \qa systems. We will describe and analyze the contribution of each module to database \qa performance, demonstrating their necessity in well-designed LLM-based database \qa systems of the future.

The workflow of the proposed testbed is shown in Figure \ref{fig:system}. 

\textbf{Offline}. Before deployment, the core LLM module goes through the stage of \textit{continual pre-training and fine-tuning} to acquire more specific DB concepts and skills while preserving the LLM's general linguistic knowledge. The user can also load a knowledge source of documents (usually up-to-date materials that do not appear in the training corpus, or in-house data for privacy reasons) stored in the form of a vectorized database.

\textbf{Online}. When the user submits a query, it first goes through the \textit{Question Classification Routing} (QCR) module to determine the logical structure of reasoning an answer. Depending on the result of QCR, i.e., the type of question, it is directed to an appropriate prompt in the \textit{Prompt Template Engineering} (PTE) module. Keywords in the prompt generated by the PTE module will trigger the \textit{Retrieval Augmented Generation} (RAG) or \textit{Tool Invocation Generation} (TIG) module to append to the content of the prompt. For example, if the query is related to a certain DB product, then to mitigate hallucination, the RAG module is triggered to retrieve trusted data and generate more accurate and relevant answers. The process can iterate for a few rounds if needed. For example, if answering the query needs to perform data analysis, the schema tool is first triggered to fetch the table structure of the database. Based on the results output by the schema tool, a selection tool is triggered to execute a SQL selection query, to compute the required data statistics in the database. Finally, the LLM is instructed by the prompt to generate the answer. 

\begin{table}[!t]
    \centering
    \caption{Comparison of LLM-based Database \qa Solutions. }
    \label{tab:framework}
\begin{small}
\vspace{-1em}
     \begin{tabular}{c|c|c|c|c|c}
        \toprule
        \thead{\textbf{Solution}} & \thead{\textbf{Pre-} \\ 
        \textbf{training}} & \thead{\textbf{Fine-} \\ \textbf{tuning}} & \thead{\textbf{Question} \\ \textbf{Routing}} & \thead{\textbf{Retrieval} \\ \textbf{Augment}} & \thead{\textbf{Tool/} \\ \textbf{Agent}} \\
        \midrule
        LLM-only & $\checkmark$ & $\checkmark$ & \textsf{X} & \textsf{X} & \textsf{X} \\
        Langchain~\cite{langchain} & \textsf{X} & \textsf{X} & \textsf{X} & $\checkmark$ & \textsf{X}$^{\ast}$ \\
        D-bot~\cite{db-gpt-zhou} & \textsf{X} & \textsf{X} & \textsf{X} & $\checkmark$ & \textsf{X}$^{\star}$ \\
        DB-GPT~\cite{db-gpt-xue} & \textsf{X} & $\checkmark$ & \textsf{X} & $\checkmark$ & \textsf{X}$^{\star}$ \\  
        \textbf{Ours} & $\checkmark$ & $\checkmark$ & $\checkmark$ & $\checkmark$ & $\checkmark$ \\
        \bottomrule
    \end{tabular}
    \smallskip
\end{small}
\parbox{0.95\linewidth}{\small \textbf{Note:} $\checkmark$: fully supports the component. \textsf{X}: lacks functionality completely. \textsf{X}$^{\ast}$: DB tools need to be customized in the Langchain framework. \textsf{X}$^{\star}$: limited support, D-bot focuses on data interaction issues, and DB-GPT focuses on database operational diagnosis.}
\vspace{-1.5em}
\end{table}

\vspace{-.5em}
\subsection{Pre-training and Fine-tuning}\label{sec:pretrain}

\textbf{(1) Pre-training}. We evaluate the effect of the continued pre-training of the backbone model on enhancing the model's expertise in the database domain. 
Specifically, we extensively collect pre-training corpora related to databases, comprising approximately 47,000 entries each in Chinese and English, totaling around 100 million tokens. This corpus includes major database textbooks, official documentation of various database products, and selected authoritative reports and articles on databases. 
For the preparation of our pre-training, we conduct a cleaning and deduplication process on the collected pre-training data. Subsequently, we process the data into text blocks containing 4096 tokens each, which are then fed into the backbone for continual pre-training. This training phase effectively enriches the model's knowledge in the database field and lays a solid foundation for subsequent fine-tuning.

\textbf{(2) Fine-tuning}. To fully evaluate the improvements in DBQA capabilities of LLMs through fine-tuning, we have proposed a customized fine-tuning strategy. Specifically, we propose a sequential fine-tuning strategy, including three stages. We prioritize the fine-tuning sequence based on the crucial abilities in DB problem-solving. For instance, the first fine-tuning stage focuses on enhancing the LLM's NL2SQL and table understanding ability using NL2SQL data like Spider~\cite{spider} because it is fundamental in DB tasks. In the second fine-tuning stage, a mixture of different fine-tuning data is adopted. The fine-tuning data includes (1) general conversational datasets like Stanford Alpaca~\cite{alpaca} to mitigate the LLM's forgetting of general dialogue skills, and (2) reformulated questions from \bench using corresponding prompts in the PTE module to enhance the LLM's understanding of the prompt template. The last fine-tuning stage focuses on enhancing the alignment of LLM's final response with DB experts in terms of quality and format, by using answer cases written by DB experts in Section~\ref{sec:bench}. The specific settings will be detailed in Section~\ref{sec:end2end}.

\subsection{Question Classification Routing}\label{sec:qcr}
The Question Classification Routing (QCR) module is designed to automatically categorize user queries and route them to different customized prompt templates. When a DBQA bot lacks the capability for categorized routing, it must rely on a single prompt template for all inputs. This limitation poses significant security risks, including the potential for inadvertently responding to legally restricted or sensitive questions. Moreover, it cannot generate accurate answers through carefully designed prompts or auxiliary modules like RAG.

In this paper, we implement and evaluate three methods of QCR modules to explore the better design paradigm.

\noindent \textbf{(1) LLM-based Classification Method.} We use a prompt, which is designed to elicit a classification response from GPT-4. \footnote{The query prompt can be found on \href{https://github.com/XMUDM/DQABench}{[link]}}

\noindent \textbf{(2) Classifier.} We train an XLNet-based~\cite{XL} classifier. We construct the training data\footnote{The sources and statistics of the dataset for these classifiers are detailed in \href{https://github.com/XMUDM/DQABench}{[link]}.} where each question is labeled as ``unsafe", ``safe but irrelevant", ``DB general", ``product-specific", or ``instance-specific". The positive samples for the ``unsafe" category are collected from Safety-Prompts~\cite{safety} and BeaverTails-Evaluation~\cite{beavertails}. The ``safe but irrelevant" samples are collected from Alpaca~\cite{alpaca} and Longbench~\cite{longbench}. The rest three categories are from the training set of \bench (which does not overlap with the test set). 

\noindent \textbf{(3) Hierarchical Classifier.} Training a single function to predict all possible labels is more difficult. Furthermore, a ``flat" classifier method requires a balanced amount of training queries for each class. Alternatively, we train a hierarchical classifier, which first classifies safe and unsafe questions and then classifies the safe questions into four sub-classes. We use an independent XLNet-based~\cite{XL} classifier at each level.

\subsection{Prompt Template Engineering} 
The Prompt Template Engineering (PTE) module includes customized prompt templates designed for various query categories. Each template incorporates slots indicated by ``\{\{\}\}," allowing for dynamic content insertion. Keywords within these templates can trigger specific modules to populate these slots with relevant data. For instance, when addressing DB product-related queries, the RAG module can be activated to supplement the prompt with retrieved external knowledge, enhancing the model's performance. Similarly, for DB instance-related queries, specific keywords can activate the TIG module to include tool-generated results within the prompt.  

Prompt engineering has consistently been a critical focus for optimizing LLMs' performance, as different prompt templates can help LLMs understand the user's intent more effectively. In this work, we standardize the templates in the testbed for general \qa, product-related \qa, instance-related \qa, and DB-irrelevant \qa to ensure a fair comparison. We try our best to refine the standardized prompt design to maximize each model's performance in the DB domain. Due to space limitations, standardized prompt templates can be found at \href{https://github.com/XMUDM/DQABench}{[link]}.

\subsection{Retrieval Augment Generation}\label{sec:RAG}
The Retrieval Augment Generation (RAG) module is used to extract additional external knowledge from documents to enhance LLMs. 

As shown in Figure~\ref{fig:system}, the module first segments texts from the knowledge base into independent text blocks. Each text block is then processed through an embedding model to be transformed into a dense vector and stored in a vector database, establishing mappings between texts and vectors. Similarly, when a user submits a query, it is also transformed into a vector using the same embedding model, which is then matched against the vectors of the text blocks in the database based on similarity computation. The system obtains the most relevant text block vector, appends it to the prompt and feeds the prompt to the core LLM module. The LLM then generates precise and relevant responses to the user’s query based on the knowledge. 

For specific RAG solutions, we follow the evaluation framework RAG-Lab~\cite{raglab} to assess six typical technical solutions on \bench as follows: \textbf{(1) Naive RAG~\cite{langchain}:} Directly utilizes the retrieved documents for generation without additional processing. \textbf{(2) RRR~\cite{rrr}:} Ranks and refines the retrieved documents to enhance the quality of responses. \textbf{(3) Iter-Retgen~\cite{iter}:} Iteratively improves both the retrieval and generation processes to achieve better output quality. \textbf{(4) Self-Ask~\cite{selfask}:} Decomposes complex queries into simpler sub-questions to facilitate retrieval. \textbf{(5) Active RAG~\cite{activerag}:} Uses active learning techniques to select the most informative documents, progressively refining the generated responses. \textbf{(6) Self-RAG~\cite{selfrag}:} A self-supervised approach that refines retrieval based on previous outputs. It is worth noting that (4), (5), and (6) all employ a sentence-by-sentence retrieval and generation approach, which incrementally refines the generated results. This strategy trades off a higher number of retrieval accesses for performance improvements in general NLP datasets.

\subsection{Tool Invocation Generation}\label{sec:TIG}
The Tool Invocation Generation (TIG) module is designed to extract context information regarding the database instance to tailor a customized answer. 

We first implement a Chain of Thought (COT) prompt template following ReAct ~\cite{react}. This prompt template encourages the LLM to think in a loop according to the following chain of thought: (1) ``Thought": Think and reason based on the currently available information to determine the tools that need to be invoked. (2) ``Action" and ``Action Input": Provide the name of the tool to be invoked and its input in the right format. (3) ``Observation": The tool will provide the results of the invocation.

As shown in Figure~\ref{fig:system}, the LLM first outputs a COT with the tool it wants to invoke and its input. The tool trigger interrupts the LLM's output, while the TIG module identifies the tool name (following ``Action:") from the pool in Table \ref{tab:tool}. If found, it calls the tool using the input specified after "Action\_Input:". Upon interacting with the database, the tool outputs results, formatted as text and appended to the LLM's output afte ``Observation:". We optimize these outputs by (1) filtering relevant content and (2) converting structured data into Markdown. After tool execution, the TIG module resumes the LLM's process, which uses the ``Observation" results to decide whether to invoke additional tools. The cycle continues until the LLM has enough information to output a final answer.

\section{End-to-End Performance} \label{sec:end2end}
We examine the end-to-end performance of different LLMs, i.e., do they produce high-quality answers for database questions. 

\subsection{Setting}
\textbf{LLMs}. Seven popular commercial and open-source LLMs are compared, including \textbf{(1) GPT-4~\cite{gpt4}}, the most powerful large-scale LLM currently released by OpenAI, using the GPT-4-0125-preview version. \textbf{(2) GPT-3.5~\cite{chatgpt}}, the most popular large-scale LLM currently released by OpenAI, using the GPT-3.5-turbo-0125 version. \textbf{(3) GLM-3~\cite{glm}}, a popular large-scale LLM for both Chinese and English, released by Zhipu AI. \textbf{(4) Llama3-8B~\cite{llama}}, the latest mid-sized open-source LLM released by Meta AI, claimed to achieve state-of-the-art (SOTA) performance among mid-sized models, using the Llama3-8B-Instruct version. \textbf{(5) Llama2-13B~\cite{llama2}}, the most popular mid-sized open-source LLM released by Meta AI, using the Llama1-13B-Chat version. \textbf{(6) Yuan2-2B~\cite{yuan}}, a popular small-sized open-source model for both Chinese and English, released by IEIT Systems, using the Yuan2-2B-Februa version. \textbf{(7) Baichuan2-13B~\cite{baichuan}}, a popular mid-sized open-source model for both Chinese and English, released by Baichuan Intelligence, using the Baichuan2-13B-Chat version. 

In addition to evaluating the performance of vanilla LLMs, we assess the impact of continued pre-training and fine-tuning for the Baichuan2-13B model, leading to two additional LLM variants: \textbf{(8) Baichuan2-13B-sft}, which is a Baichuan2-13B variant fine-tuned as shown in Section \ref{sec:pretrain}. \textbf{(9) Baichuan2-13B-cpt-sft}, which is a Baichuan2-13B variant pre-trained and fine-tuned (see Section \ref{sec:pretrain}). 


\textbf{Evaluation pipeline}. 
Directly using the LLMs can hardly generate satisfying answers for DB questions in \bench, even with prompt engineering. The workflow in Section~\ref{sec:framework}, i.e., with auxiliary techniques such as question classification routing (QCR), retrieval-augmented-generation (RAG), and tool-invocation-generation (TIG), can improve the answer quality of each tested LLM. Thus, this section presents the results generated by LLMs through the entire workflow implementing all modules of \testbed. LLM performances without the auxiliary techniques will be further investigated in Section~\ref{sec: modular}. Since each module can adopt various strategies that may introduce bias when comparing the performance between the LLMs, we use a standard pipeline with fixed intermediate output to obtain the best output of different LLMs. The details of the standard pipeline are as follows. 

\textit{(1) Question Routing with Ground-truth Class Label}. The accuracy of question classification will affect the end performance of answer generation. For example, correctly associating the question ``Why is SQL query execution slow?" with the `instance-specific' label is essential to trigger the DB tools and produce targeted answers. We will show the impact of different question classification strategies in Section~\ref{sec: modular}. In this section, to focus on the core LLM and produce the best possible answer, we use the ground-truth question labels to generate the associated prompt template.

\textit{(2) Generation with Ground-truth Retrieved Knowledge}. For product-specific questions, the retrieval documents provide the LLMs with product information. We will later show in Section~\ref{sec: modular} that the retrieval precision affects the answer quality. Thus, in this section, we use the ground truth fine-grained retrieval text (i.e., the correct text block) and append it to the prompt.

\textit{(3) Generation with Ground-truth Tool Output}. For instance-specific questions, the end-to-end evaluation focuses on LLM's ability to plan and utilize DB tools. The problem is that the tool's output may deviate from the ground truth, and the LLM will produce unexpected output, making it difficult to evaluate the personalized answer. For example, For example, LLMs may create tools or API interfaces that do not actually exist due to hallucinations. Thus, for instance-specific queries, we implement a process called "Thought-Action-Action\_Input-Observation." For each question being evaluated, the LLM is provided with a tool pool. The tool pool is initialized by including all the tools mentioned in the question's ground-truth tool labels and four randomly selected tools from the benchmark. For each tool used in the ground truth, a new prompt is generated using the prompt template and the previous tool’s output. This prompt is then passed to the LLM to decide its next action. If the LLM chooses the correct tool, the corresponding observation from the ground truth is added to its response. If it selects the wrong tool, it ends with a “Tool Invocation Failure” message.

\textbf{Testing questions}. The testing questions consist of \textit{(1) DB general questions} are fundamental concepts in the database domain, \textit{(2) product-specific questions} are about the database product `openGauss', where the external retrieval documents are openGauss latest documentation as of April 2024. The advantage of 'openGauss' is that our evaluated LLMs have not shown any signs of being specifically trained on the latest detailed documentation of this product, effectively avoiding unfair evaluation due to potential data leakage. \textit{(3) instance-specific questions} are created on the widely recognized database benchmarks TPC-H, TPC-C, and TPC-DS.

\textbf{Metrics}
We adopt two evaluation metrics, WinRate and MCA (Multiple Choice Accuracy), to measure the quality of end-to-end answer generation. 

 \textit{(1) WinRate}. This metric is widely adopted in the NLP community to score the generated answer without relying on a ground-truth answer. We compare the quality of two different LLMs, i.e., one is the LLM to be evaluated, and the other is a baseline. we use two baselines based on the most common LLM, GPT-3.5, more details are shown in Section~\ref{sec:exp_main}.  A powerful adjudicator model, GPT-4, is used to judge the quality of answers. To alleviate the length bias, i.e., the adjudicator prefers long answers, we design prompts that ask the adjudicator to focus on the basic facts of the answer rather than the language style~\footnote{Detailed prompts for WinRate, can be found at \href{https://github.com/XMUDM/DQABench}{[link]}.}. We calculate WinRate as follows:
\begin{small}
\begin{equation}
WinRate = \frac{N_{r=1}}{N_{r=1} + N_{r=-1}}, \quad \text{where} \left\{
\begin{array}{ll}
r = 1 & \text{if } M \text{ wins}, \\
r = 0 & \text{if } M \text{ ties}, \\
r = -1 & \text{if } M \text{ loses},
\end{array}
\right. 
\end{equation}
\end{small} where $N_r$ represents the number of comparisons where the judge GPT-4 considers case $r$, and $M$ is the model to be evaluated.

\textit{(2) MCA (Multiple-Choice Accuracy)}. To complement the objective LLM-based evaluation WinRate, we also provide subjective evaluations on DB-general questions. This metric measures the accuracy of all multiple-choice questions following $MCA =\sum_{i} m_{i,i}/\sum_{i}$ $ \sum_{j} m_{i,j}),$ where $m_{i,j}$ is the number of answers that the ground truth choice is $i$ and the LLM's output is $j$, $i,j \in \{A, B, C, D, others\}$. For MCA, we prompt the LLM to output one letter; any deviation is classified as $others$ and considered a categorization error.

\textbf{Other implementation details}. The LLMs are trained on a workstation equipped with eight A100 GPU cards. The pre-training phase comprises two epochs using the collected database-related data. The learning rate is 5e-5, and the batch size is 128. In the fine-tuning phase, the four subsets, namely General DB Q\&A, Product-specific Q\&A, Instance-specific Q\&A, and DB-irrelevant Q\&A, are assigned with weights 1:1.5:3:0.5, respectively, to account for their importance, data volume, and training difficulty. The learning rate is 2e-5, and the batch size is 64.

\vspace{-.5em}
\subsection{Main Result} \label{sec:exp_main}
\begin{table*}[t]
\centering
\setlength{\tabcolsep}{2pt}
\caption{WinRate of different LLMs versus the competitor}
\label{tab:main_WinRate_Result}
\vspace{-1em}
\resizebox{0.90\linewidth}{0.30\linewidth}{
\begin{tabular}{lrr|cc|cc|cc|cc}
\toprule
\multirow{2}{*}{\textbf{Model}} & \multirow{2}{*}{$|\boldsymbol{\theta}|$} & \multirow{2}{*}{\textbf{Deployment}} & \multicolumn{2}{c|}{\textbf{DB General}} & \multicolumn{2}{c|}{\textbf{Product-specific}} & \multicolumn{2}{c|}{\textbf{Instance-specific}} & \multicolumn{2}{c}{\textbf{Average}}\\
\cmidrule{4-11}
& & & ZH & EN & ZH & EN & ZH & EN & ZH & EN \\
\midrule
\multicolumn{10}{c}{\texttt{WinRate v.s. Vanilla GPT-3.5-Turbo}} \\
\midrule
\href{https://platform.openai.com/docs/models/gpt-4-turbo-and-gpt-4}{GPT-4} &  & Centralized & \hl{0.85} & \hl{0.95} & 0.86 & \hl{0.86} & 0.69 & 0.53 & \hl{\textbf{0.80}} & \hl{\textbf{0.78}}\\
\href{https://platform.openai.com/docs/models/gpt-3-5-turbo}{GPT-3.5-Turbo} & & Centralized & 0.53 & 0.56 & 0.60 & 0.60 & 0.58 & 0.57 & \textbf{0.57} & \textbf{0.58}\\
\href{https://chatglm.cn/main/detail}{GLM-3-Turbo} & & Centralized & 0.63 & 0.62 & 0.81 & 0.58 & 0.44 & 0.44 & \textbf{0.63} & \textbf{0.55}\\
\href{https://huggingface.co/meta-llama/Meta-Llama-3-8B-Instruct}{Llama3} & 8B & $\geq$RTX 3090 & 0.60 & 0.67 & 0.79 & 0.75 & 0.37 & 0.40 & \textbf{0.59} & \textbf{0.61}\\
\href{https://huggingface.co/meta-llama/Llama-2-13b-chat-hf}{Llama2} & 13B & $\geq$RTX 4090 & 0.12 & 0.09 & 0.35 & 0.41 & 0 & 0 & \textbf{0.16} & \textbf{0.17}\\
\href{https://huggingface.co/IEITYuan/Yuan2-2B-Februa-hf}{Yuan2} & 2B & $\geq$RTX 3060 & 0.03 & 0.02 & 0.22 & 0.18 & 0 & 0 & \textbf{0.08} & \textbf{0.07}\\
\href{https://huggingface.co/baichuan-inc/Baichuan2-13B-Chat}{Baichuan2}(vanilla) & 13B & $\geq$RTX 4090 & 0.27 & 0.29 & 0.56 & 0.60 & 0.23 & 0.15 & \textbf{0.35} & \textbf{0.35}\\
Baichuan2-sft & 13B & $\geq$RTX 4090 & 0.44 & 0.31 & 0.88 & 0.76 & 0.90 & 0.82 & \textbf{0.74} & \textbf{0.63} \\
Imp. w.r.t. vanilla & & & \grey{+0.17} & \grey{+0.02} & \grey{+0.32} & \grey{+0.16} & \grey{+0.67} & \grey{+0.67}& \grey{+0.39} & \grey{+0.28}\\
Baichuan2-cpt-sft & 13B & $\geq$RTX 4090 & 0.57 & 0.48 & \hl{0.88} & 0.74 & \hl{0.91} & \hl{0.87} & \textbf{0.79} & \textbf{0.70}\\
Imp. w.r.t. vanilla & & & \grey{+0.30} & \grey{+0.19} & \grey{+0.32} & \grey{+0.14}  & \grey{+0.68} & \grey{+0.72}& \grey{+0.44} & \grey{+0.35}\\
\midrule
\multicolumn{10}{c}{\texttt{WinRate v.s. GPT-3.5-Turbo (Testbed)}} \\
\midrule
\href{https://platform.openai.com/docs/models/gpt-4-turbo-and-gpt-4}{GPT-4} & & Centralized  & \hl{0.83} & \hl{0.95} & 0.64 & 0.68 & 0.90 & 0.64 & \hl{\textbf{0.79}} & \hl{\textbf{0.76}}\\
\href{https://platform.openai.com/docs/models/gpt-3-5-turbo}{GPT-3.5-Turbo} & &Centralized &- & - & - & - & - & - & - & -\\
\href{https://chatglm.cn/main/detail}{GLM-3-Turbo} & &Centralized & 0.62 & 0.65 & 0.66 & 0.57 & 0.55 & 0.49 &\textbf{0.61} & \textbf{0.57}\\
\href{https://huggingface.co/meta-llama/Meta-Llama-3-8B-Instruct}{Llama3} & 8B & $\geq$RTX 3090 & 0.60 & 0.65 & 0.62  & 0.51 & 0.49 & 0.52 & \textbf{0.57} & \textbf{0.56}\\ 
\href{https://huggingface.co/meta-llama/Llama-2-13b-chat-hf}{Llama2} & 13B & $\geq$RTX 4090 & 0.12 & 0.06 & 0.36 & 0.16 & 0 & 0 & \textbf{0.16} & \textbf{0.07}\\
\href{https://huggingface.co/IEITYuan/Yuan2-2B-Februa-hf}{Yuan2} & 2B & $\geq$RTX 3060 & 0.03 & 0.02 & 0.13 & 0.07 & 0 & 0 & \textbf{0.05} & \textbf{0.03}\\
\href{https://huggingface.co/baichuan-inc/Baichuan2-13B-Chat}{Baichuan2(vanilla)} & 13B & $\geq$RTX 4090 & 0.26 & 0.30 & 0.42 & 0.40 & 0.16 & 0.11 & \textbf{0.28} & \textbf{0.27}\\
Baichuan2-sft & 13B & $\geq$RTX 4090  & 0.44 & 0.30 & \hl{0.68} & 0.66 & 0.85 & 0.87 & \textbf{0.66} & \textbf{0.61} \\
Imp. w.r.t. vanilla & & & \grey{+0.18} & \grey{0} & \grey{+0.26} & \grey{+0.26} & \grey{+0.69} & \grey{+0.76} & \grey{+0.38} & \grey{+0.34} \\
Baichuan2-cpt-sft & 13B & $\geq$RTX 4090 & 0.55 & 0.42 & 0.65 & \hl{0.73} & \hl{0.95} & \hl{0.90} & \textbf{0.72} & \textbf{0.68} \\
Imp. w.r.t. vanilla & & & \grey{+0.29} & \grey{+0.12} &\grey{+0.23} & \grey{+0.33} & \grey{+0.79} & \grey{+0.79}& \grey{+0.44} & \grey{+0.41}\\
\bottomrule
\end{tabular}
}
\vspace{-1em}
\end{table*}

We compute the WinRate of each LLM versus two baselines. \textit{(1) GPT-3.5-Turbo (Vanilla)}: we input the user's question to GPT-3.5-Turbo without employing any prompts. This result demonstrates the difference between a dedicated DBQA system (i.e., the testbed) and typical LLM applications. \textit{(2) GPT-3.5-Turbo (Testbed)}: we also equip GPT-3.5-Turbo on the testbed. This result emphasizes the inherent capability difference of each LLM. We have the following insights from Table~\ref{tab:main_WinRate_Result}.

\noindent \textbf{I1: Larger model sizes and richer pre-training data improve answer qualities on DB general questions. } GPT-4 has achieved SOTA performance on DB general questions. Generally, among LLMs released in the same period, larger model sizes tend to perform better. However, Llama3-8B challenges this trend. As stated in its technical report, Llama3-8B introduced an extensive amount of training data (15 trillion tokens) during pre-training, resulting in logarithmic-scale performance improvements. This enhancement is also evident on the \bench dataset.

\noindent \textbf{I2: Domain-specific continual pre-training and fine-tuning can significantly improve DBQA performance}. The testbed's continual pre-training and fine-tuning have increased the performance of Baichuan2-13B on all question types. Specifically, the pretraining has increased the average performance by 0.28 to 0.39 (80\% to 111\%). The fine-tuning has significantly improved answers to all questions, with an average overall increase of 0.28 to 0.38 (80\% to 136\%). Notably, the fine-tuning stage targets to reinforce instruction adherence to specific prompt templates and gives the LLM a stronger ability to align with the input format of various DB tools. Consequently, we observe that Baichuan2-13B-sft outperforms GPT-4 on the instance-specific questions.

\noindent \textbf{I3: RAG and TIG on the testbed can enhance LLM performance in DBQA tasks.} Comparing the upper half with the bottom half of Table~\ref{tab:main_WinRate_Result}, we observe that the WinRate has generally decreased because GPT-3.5-Turbo on the testbed is more competitive than GPT-3.5-Turbo vanilla. It shows that, even for a sophisticated large-scale LLM, the RAG and TIG on the testbed can be beneficial. We will discuss the impacts of RAG and TIG in more detail in Section \ref{sec:exp_rag} and \ref{sec:exp_tig}.

\noindent \textbf{I4: Small-sized general-purpose LLMs can hardly invoke DB tools}. In answering instance-specific questions, the DB tools' input and output often deviate from standard templates, and the LLM's ability to follow the question's instructions is fundamental to meeting specific tool usage requirements. Among all the tested LLMs, Llama2 and Yuan2 are relatively smaller in model size and have not been sufficiently pre-trained for instruction-following. As a result, they exhibit poor tool usage capabilities. Although models like Baichuan2-13B claim to have conducted instruction-following alignment, experimental results show that their model size still significantly limits their tool usage capabilities. This indicates that instruction enhancement for specific tool usage is necessary when deploying small to medium-sized models. We will further discuss the accuracy of the TIG module in Section \ref{sec:exp_tig}.

\subsection{In-depth Analysis on DB General Questions}\label{sec:exp_general}
The DB-general questions require the LLM to depend on its inherent knowledge to answer the questions, where each question may require a different skill set. For example, the LLM needs to be proficient in SQL grammar to solve ``Write a SQL to create index" and knowledge of index advisor to answer ``which index is better."  There are two types of questions in the subset, namely the subjective questions and the objective questions with multiple choices.  In this section, we analyze the answers to DB-general questions from the two perspectives. 

First, we report the WinRate of various LLMs in answering subjective questions. Specifically, we utilize GPT-4 to assign detailed labels to each question in the ``DB general'' subset, incorporating predefined few-shot labels from the prompt and allowing GPT-4 to generate new labels autonomously~\footnote{The details of the classification prompt can be found at \href{https://github.com/XMUDM/DQABench}{[link]}}. We identify the eight most common labels: ``Performance Monitoring and Tuning'', ``Backup and Recovery'', ``Query Optimization'', ``Data Migration'', ``Data Security'', ``Database Design and Deployment'', ``Data Analysis'' and ``SQL Programming''. These labels cover 93.79\% of the questions in the ``DB general'' subset.

\begin{figure}[!t]
  \centering
\includegraphics[width=\linewidth]{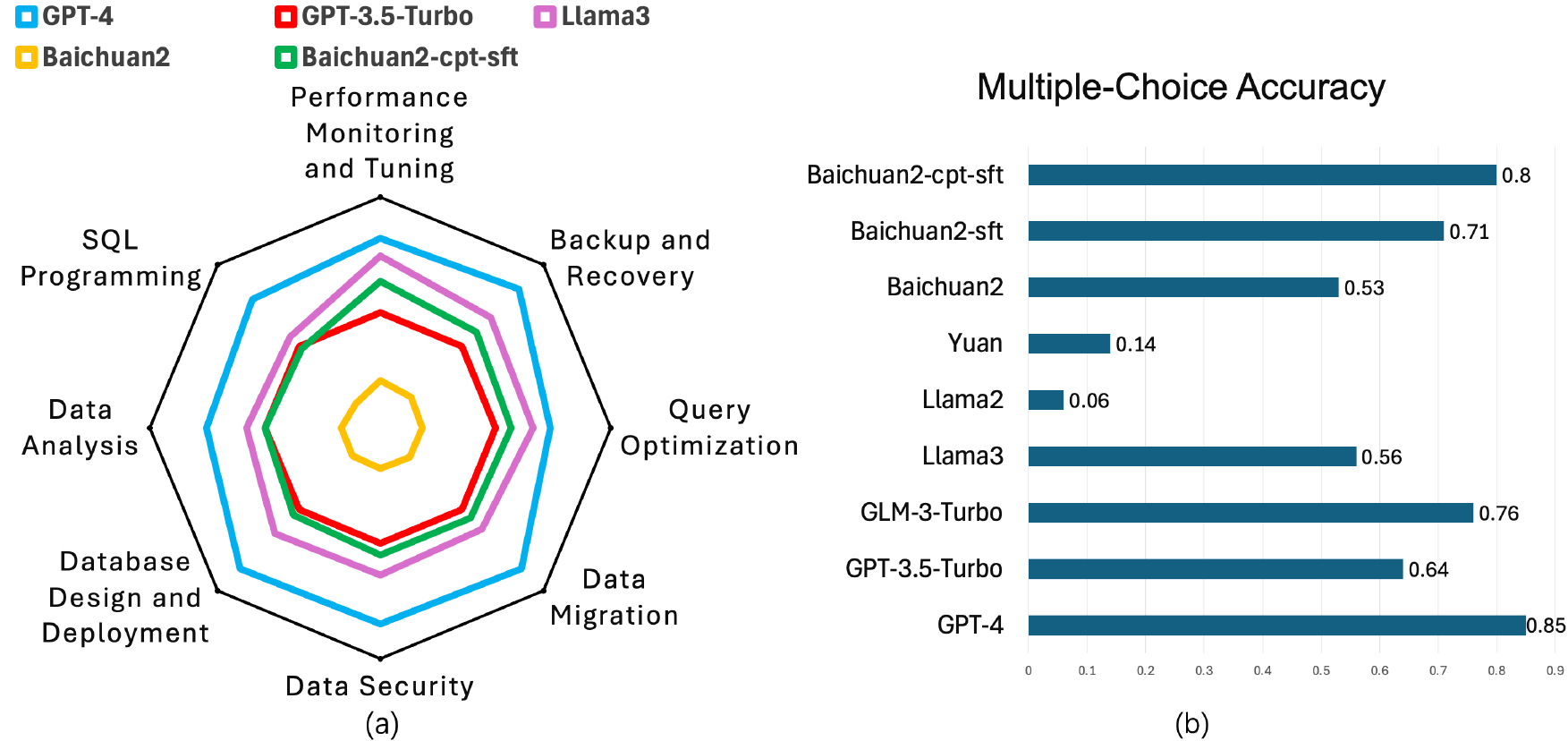}
  \vspace{-1em}
  \caption{(a) WinRate and (b) Multi-Choice Accuracy on ``DB General" questions}
  \label{fig:exp_general}
  \vspace{-1em}
\end{figure}

The experimental results, illustrated in Figure~\ref{fig:exp_general} (a)'s radar chart, lead to the following conclusions: (1) Numerically, the response capabilities of different models in each sub-field correlate positively with their model size and overall ability, with no model being highly specialized in any particular field. (2) Regarding shape proportion, the radar charts of GPT-3.5 and GPT-4 are similar, showing balanced capabilities across the eight aspects. In contrast, Llama2, Llama3, Baichuan2, and other models based on the llama architecture display a similar pattern, excelling in Performance Monitoring and Tuning but weaker in SQL Programming. This indicates that GPT series models are better at generating accurate SQL Programming instructions, while llama-based models excel in comprehensive subjective analysis. 

Next, we report the Multiple-Choice Accuracy, measured on the objective questions. As shown in Figure~\ref{fig:exp_general} (b), MCA aligns closely with the WinRate on subjective questions. This validates that questions in \bench are set with an appropriate difficulty level and require a good understanding of DB knowledge to answer.
\section{Modularized Evaluation} \label{sec: modular}
In this section, we rigorously assess the performance of each module of the testbed \testbed, including question classification routing (QCR), retrieval-augmented-generation (RAG), and tool-invocation-generation (TIG). We aim to achieve two goals: (1) verify the necessity of adopting each module for DBQA to improve answer quality; (2) evaluate the strengths and weaknesses of various solutions for each module in the context of DBQA.

\subsection{Experimental Setup}
\textit{(1) QCR:} For all Chinese classifiers, we train them using XLNet-base~\cite{XL}, set the batch size to 512 and the learning rate to 5e-4. For all English classifiers, we train them with XLNet-base~\cite{XL} using a batch size of 256 and a learning rate of 5e-5. Each classifier is trained for 30 epochs. Checkpoints that perform best on the validation sets are saved. \textit{(2) RAG:} the text length is set to 250 characters per segment, and the overlap length is 50 characters between adjacent texts in the knowledge base. For each query, up to three vectors from the knowledge base are matched with a similarity threshold of 0.5. We use L2 distance for similarity search with the default Flat index. \textit{(3)TIG:} We set the number of random tools in the tool pool ({{T}} in prompt) to 4.

\subsection{Modularized Evaluation on QCR}

\begin{table}[t!]
\centering
\caption{WinRate (w/ Routing v.s. w/o Routing)}
\vspace{-1em}
\begin{tabular}{c|cc|cc}
    \hline
    \multirow{3}{*}{\textbf{Models}} & \multicolumn{4}{c}{\textbf{WinRate V.S. Self w/o Routing}} \\ 
    \cline{2-5} & \multicolumn{2}{c|}{\textbf{Product-Specific}} & \multicolumn{2}{c}{\textbf{Instance-Specific}} \\
    \cline{2-5} & \textbf{ZH} & \textbf{EN} & \textbf{ZH} & \textbf{EN} \\
    \midrule
GPT-4              &        0.85        &           0.75          & 0.69           & 0.61                \\
GPT-3.5-turbo      &        0.73        &           0.76          & 0.59           & 0.60                \\
GLM-3-turbo        &        0.82       &           0.76          & 0.44           & 0.53                \\
Llama3-8B-Instruct &         0.84       &           0.78          & 0.41           & 0.60                \\
Llama2-13B-Chat    &        0.90        &           0.81          & 0              & 0                   \\
Yuan-2B            &        0.96        &           0.94           & 0              & 0                   \\
Baichuan2-13B-Chat &        0.68        &           0.73        & 0.26           & 0.07                \\
Baichuan2-sft      &         0.52       &           0.65        & 0.87           & 0.82                \\
Baichuan2-cpt-sft  &         0.56       &           0.73          & 0.86           & 0.86  \\      
\bottomrule       
\end{tabular}
\vspace{-1em}
\label{tab:winrate_self_comparison}
\end{table}

We implement two versions of each LLM to test whether question routing is necessary for  DBQA systems. (1) \textit{LLM w/ Routing}: using the QCR model for question classification and feeding the LLM with customized prompts according to the question type. We use the ground truth question type in \bench. (2) \textit{LLM w/o Routing}:  prompting the LLM using the "General DB" prompt template for all questions. Table \ref{tab:winrate_self_comparison} reports the WinRate of LLM w/Routing v.s. LLM w/o Routing on two question types, i.e., Product-Specific and Instance-Specific, because DB General questions are treated using the same prompt and thus receive the same result.

The results indicate that: (1) Most WinRate values exceed 0.5 (meaning LLM w/ Routing surpasses LLM w/o Routing), which suggests that utilizing query routing to customize responses based on question type significantly enhances the LLM's performance. Recent studies have demonstrated the importance of prompt engineering~\cite{COT}. The QCR module can be seen as a dedicated, prompt engineering strategy tailored for DBQA that autonomously gives instructions and organizes examples. (2) In a few cases, particularly on instance-specific questions, the WinRate values fall below 0.5 or even reach zero. This is due to the model's limited ability to invoke tools: models lacking any tool invocation capability cannot provide effective customized responses, and thus, QCR can not improve their performance.

Having verified the benefits of question routing based on question type in DBQA, we then investigate which question classifier in Section~\ref{sec:framework} is effective. The testing data include DB-related questions from \bench, safe but DB-irrelevant questions from Alpaca~\cite{alpaca} and Longbench~\cite{longbench} and unsafe labeled questions from Safety-Prompts~\cite{safety} and BeaverTails Evaluation~\cite{beavertails}~\footnote{The sources and statistics of the dataset are detailed in \href{https://github.com/XMUDM/DQABench}{[link]}.}. The F1 score for each category, accuracy, and response latency deployed on a workstation with an RTX-3090 GPU card are shown in Table~\ref{exp_qcr}.

\setlength{\tabcolsep}{0.9pt}
\begin{table}[t!]
\centering
\caption{Classification Performance}
\label{exp_qcr}
\vspace{-1em}
\begin{tabular}{l|ccccc|c|c}
\toprule
\multirow{2}{*}{\textbf{Method}} & \ & & \textbf{F1-score} & &  & \multirow{2}{*}{\textbf{ACC}} & \multirow{2}{*}{\textbf{Latency}} \\
& \textbf{Unsafe} & \textbf{General} & \textbf{Gauss} & \textbf{Tool} & \textbf{No-DB} & &  \\
\midrule
 \multicolumn{8}{c}{\textbf{ZH}} \\
\midrule
GPT-4  & 0.77 & 0.48 & 0.47 & 0.25 & 0.59 & 0.55 &  2.64s \\
XLNet  & 0.95 & 0.89 & 0.92 & 0.91 & 0.79 & 0.90 & \hl{0.39s} \\
Hierarchical  & 0.95 & 0.91 & 0.98 & 0.99 & 0.87 & \hl{0.94} & 0.68s \\
\midrule
 \multicolumn{8}{c}{\textbf{EN}} \\
\midrule
GPT-4 & 0.80 & 0.57 & 0.37 & 0.49 & 0.69  & 0.63 & 2.61s \\
XLNet  & 0.91 & 0.92 & 0.81 & 0.90 & 0.79 & 0.87 & \hl{0.37s} \\
Hierarchical & 0.92 & 0.97 & 0.88 & 0.93 & 0.91 & \hl{0.92}  & 0.67s \\
\bottomrule
\end{tabular}
\vspace{-1em}
\end{table}

From the experimental results, we have the following conclusions: (1) XLNet and Hierarchical are more accurate (+0.35) and faster (6.7x) than GPT-4. This is reasonable because general-purpose LLMs like GPT without fine-tuning perform worse on specific tasks than smaller models that are specifically trained. (2) There is a trade-off between latency and accuracy for small models. As shown in Table~\ref{exp_qcr}, the hierarchical classifier can achieve approximately a 0.05 performance improvement with 300ms more inference cost.

\subsection{Modularized Evaluation on RAG}\label{sec:exp_rag}
We have demonstrated the importance of RAG in end-to-end DBQA in Section~\ref{sec:exp_main} by the sharp performance rise with RAG. In Section~\ref{sec:exp_main}, GPT-3.5-Turbo is provided the ground-truth retrieval text blocks. In this section, we want to investigate whether RAG with incorrect retrieval results can enhance answer generation. 

\begin{figure}[!t]
  \centering
\includegraphics[width=0.97\linewidth]{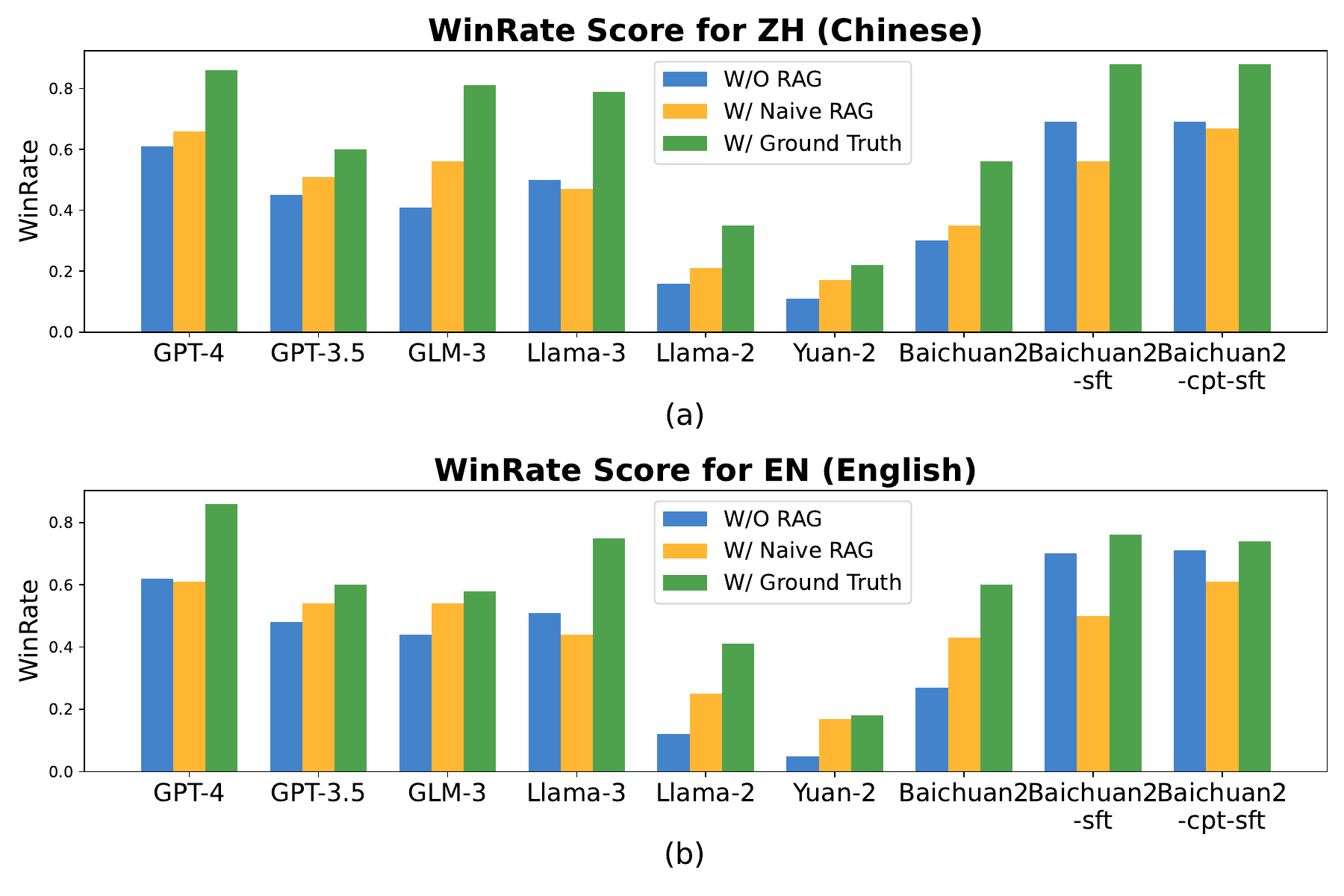}
  \vspace{-2em}
  \caption{WinRate v.s. GPT-3.5-Turbo (vanilla)}
  \label{fig:exp_rag}
  \vspace{-.5em}
\end{figure}

First, we implement three versions of the nine LLMs to be evaluated. (1) w/o RAG, the LLMs are prompted to generate answers without the external knowledge provided by the RAG module; (2) w/ Naive RAG, the testbed retrieves the relevant text blocks by directly searching the vector database; (3) w/ Ground-truth RAG, the testbed uses the retrieval ground-truth from \bench. We report the WinRate score of each LLM version v.s. GPT-3.5-Turbo (vanilla) on the ``product-specific" sub-dataset in Figure \ref{fig:exp_rag}.

From the results, we have the following observations. (1) If the ground-truth retrieval texts are provided, the RAG module can significantly enhance the performance of general-purposed LLMs of any size, i.e., the seven LLMs. The performance improvements are more pronounced in smaller models, e.g.,  a $2.4\times$ improvement on Llama2. This observation reveals the substantial value of RAG for edge-deployed models. However, the improvement is under ideal conditions, while the actual retrieval accuracy is not guaranteed in real applications. (2) The ground-truth RAG implementation has brought minimal performance improvement for models fine-tuned with domain knowledge from databases, i.e., the last two LLMs Baichuan2-sft, and Baichuan2-cpt-sft. This observation suggests a significant overlap between the improvements brought by model fine-tuning and those from the RAG module, indicating that deploying either technique alone is sufficient within a limited budget. (3) With the naive RAG implementation, the performance increase reasonably shrinks for most LLMs. Furthermore, the performance improvement is negligible for GPT-4, and we observe a performance decline on Llama-3, Baichuan2-sft, and Baichuan2-cpt-sft. Detailed case studies indicate that this is primarily due to low recall rates. LLMs generate incorrect responses by grounding on irrelevant documents, ultimately compromising their ability to produce high-quality answers for questions they could have answered otherwise correctly without RAG. 

\begin{figure}[!t]
  \centering
\includegraphics[width=0.97\linewidth]{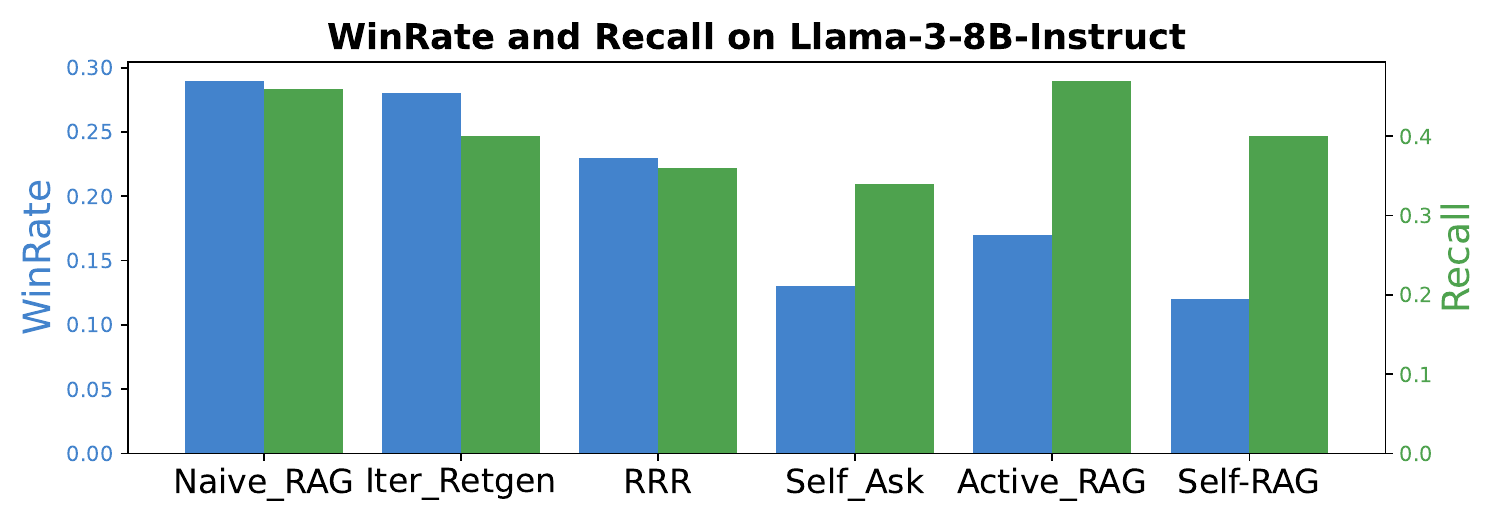}
  \vspace{-1.5em}
  \caption{WinRate and Recall Rate for Different RAG Solutions with Llama-3-8B-Instruct}
  \label{fig:exp_rag2}
\end{figure}

Second, we investigate the impact of various RAG techniques implemented in \testbed. We report two metrics in Figure~\ref{fig:exp_rag2}: (1) \textit{Recall rate}: the average number of relevant text blocks successfully identified in the top-3 retrieval results, where the ground-truth labels contain one relevant block for each question. Since the result may not be completely identical with the ground-truth text block, i.e., they have overlapping parts. To determine whether a result is relevant, we compute the ROUGE-5, i.e., the overlap of sequences of five consecutive characters between the result and the ground-truth. If \texttt{ROUGE-5 > 0.15}, we determine a result is relevant. (2) \textit{WinRate} calculates the performance of Llama3-8B-Instruct with the aformentioned RAG methods, compared with GPT-3.5-Turbo. 

We have the following observations. (1) All existing RAG techniques yield low recall rates (below 50\%), highlighting the main challenge in DBQA: accurately locating relevant documents. (2) For RAG methods that generate a passage based on the retrieval results, higher retrieval performance generally leads to higher answer quality, i.e., the recall rate positively correlates with the WinRate on Naive\_RAG, Iter\_Retgen, and RRR. (3) Techniques using a sentence-by-sentence generation strategy, such as Self-Ask, Active RAG, and Self-RAG, harm the QA performance. Because these RAG strategies divide the answer passage into sentences and generate sentences based on relevant documents at the sentence level, their retrieval module has been called more often, which raises the risk of encountering irrelevant information. For example, we identify an erroneous case where the question concerns only the \texttt{log\_dir} parameter, but the retrieval results also contain information on the \texttt{alarm\_report} parameter, the LLM elaborates on both parameters in the answer, which is undesired.

\subsection{Modularized Evaluation on TIG}\label{sec:exp_tig}
The TIG (Tool Invocation Generation) module enhances the ability of LLMs to interact with database systems, particularly in DB instance-related \qa. The benefit of TIG is already highlighted in the end-to-end experiment, i.e., \testbed improves GPT-3.5-Turbo on Instance-specific questions in Table~\ref{tab:main_WinRate_Result}. Specifically, we examine the correlation between win rate and tool invocation success rate. Our findings show that nearly all (approximately 93\%) responses, where the tool invocations are successful,  achieve a win in the final answer comparison. 
In this section, we further explore the capabilities of existing LLMs and discuss how LLMs can support more effective instance-related Q\&A.

\noindent \textbf{Metrics}. We propose TSA (Tool Selection Accuracy) and TFA (Tool Format Accuracy) to measure tool invocation capabilities.

\textit{(1) TSA} measures whether the correct tools are chosen to solve problems. It is essential to consider the order of actions to measure the DB-specific planning ability. For example, the input of the succeeding tool is usually based on and formulated from the preceding tool's output. Therefore, if there is an error in the current tool invocation, the subsequent invoked tools will no longer be included in the metric calculation. Specifically, the TSA (Tool Selection Accuracy) is defined as:
$TSA = \sum_{1\leq i\leq \min{k_j, I\{t_{k_j,j}\}=0},j}I\{t_{i,j},j\}\big{/}\sum_j k_j$, where $t_{i,j}$ is $i-$th tool for the query $j$, $I\{\}$ is an indicator function that returns whether the tool (the name after ``Action") is labeled in the tool annotation in Section~\ref{sec:bench}, and $k_j$ means the number of LLM tool invocations. 

\textit{(2) TFA} measures the accuracy of the tool invocation format, i.e., whether the LLM's response aligns with the tool's input. Due to the diversity and subjectivity of tool format requirements, particularly in the generalized tool \qa, it is challenging to assess format compliance using predefined rules. Therefore, we employ GPT-4 as an expert adjudicator model to judge whether tool invocations meet the format requirements. Similarly, we consider the order of tools. 
Specifically, the TFA (Tool Format Accuracy) is defined as:
$TFA = \sum_{1\leq i\leq \min{k_j, G\{t_{k_j,j}\}=0},j}G\{t_{i,j}\}\big{/}\sum_j k_j$, where $t_{i,j}$ is the $i-$th tool for the query $j$, $G\{\}$ is the output of GPT-4 that decides whether the tool input (the content after ``Action\_Input") is correct, and $k_j$ means the number of LLM tool invocations. 

\begin{figure}[!t]
  \centering
\includegraphics[width=\linewidth]{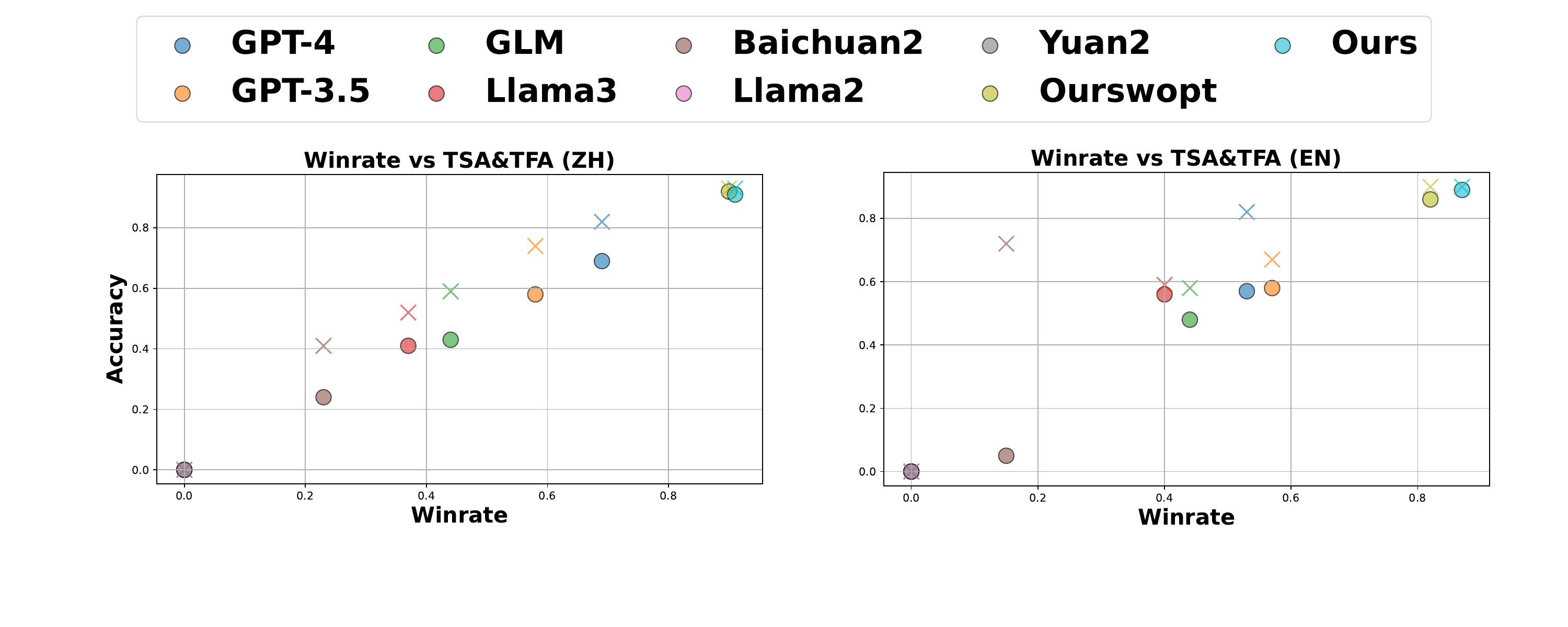}
  \vspace{-2em}
  \caption{TFA and TSA vs. Performance of Different LLMs}
  \label{fig:tool}
  \vspace{-2em}
\end{figure}

\noindent \textbf{Experimental Result.} The experimental results are shown in Figure~\ref{fig:tool}. From these results, we observe the following: (1) The ability of LLMs to select appropriate tools and format input correctly is closely correlated with their overall performance. (2) There is a significant difference in the ability of the tested LLMs to invoke database tools. Comparing Figure~\ref{fig:tool} and Table 5, we observe that the difference in tool invocation ability is even greater than in answering general or product-specific questions. For instance, the WinRate of llama2 and yuan in TIG is close to zero, while the performance of baichuan2 variants approaches one. (3) The capability of LLMs to invoke database tools is impacted by whether the models underwent instruction-following fine-tuning, i.e., exposing the LLMs to various examples where they learn to interpret and respond to instructions in a particular format or style, and alignment, i.e., adjusting the LLMs to align their behaviors with certain goals. Specifically, the technical reports for LLaMA2~\cite{llama} and Yuan~\cite{yuan} indicate that they lack targeted instruction-following fine-tuning and alignment, producing the worst  TSA and TFA results. In contrast, in addition to the general instruction-following fine-tuning of Baichuan-13B backbone, \testbed implements fine-tuning on DB-related examples to enhance Baichuan-13B's ability to follow DB-specific instructions and obtains the best TSA and TFA results.

\section{Conclusion}
In this paper, we proposed the first comprehensive DBQA benchmark \bench. First, we proposed a comprehensive dataset, which includes an extensive \qa dataset and corresponding generation methods. Second, we proposed a complete testbed that implements the entire DBQA workflow, incorporating various auxiliary modules for DB \qa. Third, we propose a complete evaluation pipeline and conducted a comprehensive evaluation to showcase DB \qa ability of seven general-purpose LLMs and two variants based on pretraining and fine-tuning. Fourth, we also assess the impact of modules such as QCR, RAG and TIG on DBQA performance and identify future directions for improvement by evaluationg existing solutions. We hope our benchmark and findings will better guide the future development of LLM-based DBQA research.

\clearpage


\end{document}